# Electronic phase separation in topological surface states of rhombohedral graphite

Yanmeng Shi[1,†], Shuigang Xu[2,†], Yaping Yang[1,2,†], Sergey Slizovskiy[2], Sergei V. Morozov[3], Seok-Kyun Son[1,2], Servet Ozdemir[1], Ciaran Mullan[1], Julien Barrier[1,2], Jun Yin[1,2], Alexei I. Berdyugin[1], Benjamin A. Piot[4], Takashi Taniguchi[5], Kenji Watanabe[5], Vladimir I. Fal'ko[1,2,6], Kostya S. Novoselov[1,2,7,8], A. K. Geim[1,2], Artem Mishchenko[1,2,*]

[1]Department of Physics and Astronomy, University of Manchester, Manchester M13 9PL, UK.
[2]National Graphene Institute, University of Manchester, Manchester M13 9PL, UK.
[3]Institute of Microelectronics Technology and High Purity Materials, Russian Academy of Sciences, Chernogolovka, 142432, Russia.
[4]Laboratoire National des Champs Magnétiques Intenses, LNCMI-CNRS-UGA-UPS-INSA-EMFL, 25 avenue des Martyrs, 38042 Grenoble, France.
[5]National Institute for Materials Science, 1-1 Namiki, Tsukuba, 305-0044, Japan.
[6]Henry Royce Institute for Advanced Materials, Manchester, M13 9PL, UK.
[7]Centre for Advanced 2D Materials, National University of Singapore, 117546, Singapore
[8]Chongqing 2D Materials Institute, Liangjiang New Area, Chongqing 400714, China
[†]These authors contributed equally
*e-mail: artem.mishchenko@gmail.com

**Of the two stable forms of graphite, hexagonal (HG) and rhombohedral (RG), the former is more common and has been studied extensively. RG is less stable, which so far precluded its detailed investigation, despite many theoretical predictions about the abundance of exotic interaction-induced physics[1-6]. Advances in van der Waals heterostructure technology[7] have now allowed us to make high-quality RG films up to 50 graphene layers thick and study their transport properties. We find that the bulk electronic states in such RG are gapped[8] and, at low temperatures, electron transport is dominated by surface states. Because of topological protection, the surface states are robust and of high quality, allowing the observation of the quantum Hall effect, where RG exhibits phase transitions between gapless semimetallic phase and gapped quantum spin Hall phase with giant Berry curvature. An energy gap can also be opened in the surface states by breaking their inversion symmetry via applying a perpendicular electric field. Moreover, in RG films thinner than 4 nm, a gap is present even without an external electric field. This spontaneous gap opening shows pronounced hysteresis and other signatures characteristic of electronic phase separation, which we attribute to emergence of strongly-correlated electronic surface states.**

Films of rhombohedral graphite (RG), also known as chirally-stacked ABC graphene multilayers, possess nearly flat bands localised at the surfaces which are particularly interesting for exploring electron-electron interactions[3]. In RG, the interlayer hopping $\gamma_1$ dimerises the electronic states of the opposite sublattices of all contiguous graphene layers in the bulk, shifting their energies away from the Fermi level. Electrons of only two sublattices, one in the top layer, and the other in the bottom layer, remain at low energies. The surface states residing on the inequivalent sublattices of the opposite surfaces, in $N$-layer-thick samples, due to $N$-layer hopping, are supposed to form nearly flat bands with approximately $E \approx \pm p^N$ dispersion relation[3,9]. These surface states are theorised to have strong electron-electron interactions, hypothetically leading to spontaneous quantum Hall states, ferromagnetism and superconductivity[1-6]. When considering other hopping parameters (Fig. 1a, also Supplementary Materials), the surface flat bands acquire a finite bandwidth of the order of $2\gamma_4\gamma_1/\gamma_0$ as well as electron-hole asymmetry, and trigonal warping due to $\gamma_2$ and $\gamma_3$, Fig. 1b, which could affect electron-electron interactions, spontaneous symmetry breaking, and the opening of a band gap, Fig. 1c. Due to the metastable nature of RG, the transport experiments are limited to tri- and four-layer ABC-stacked graphene, either in suspended two-terminal devices, or with the spectrum modified by moiré superlattices[10-12]. Here we report electronic transport in pristine high-quality RG films of up to ≈50 graphene layers thick, where we observe strong electronic correlations and phase separation typically reserved for unconventional superconductors and heavy fermion materials.

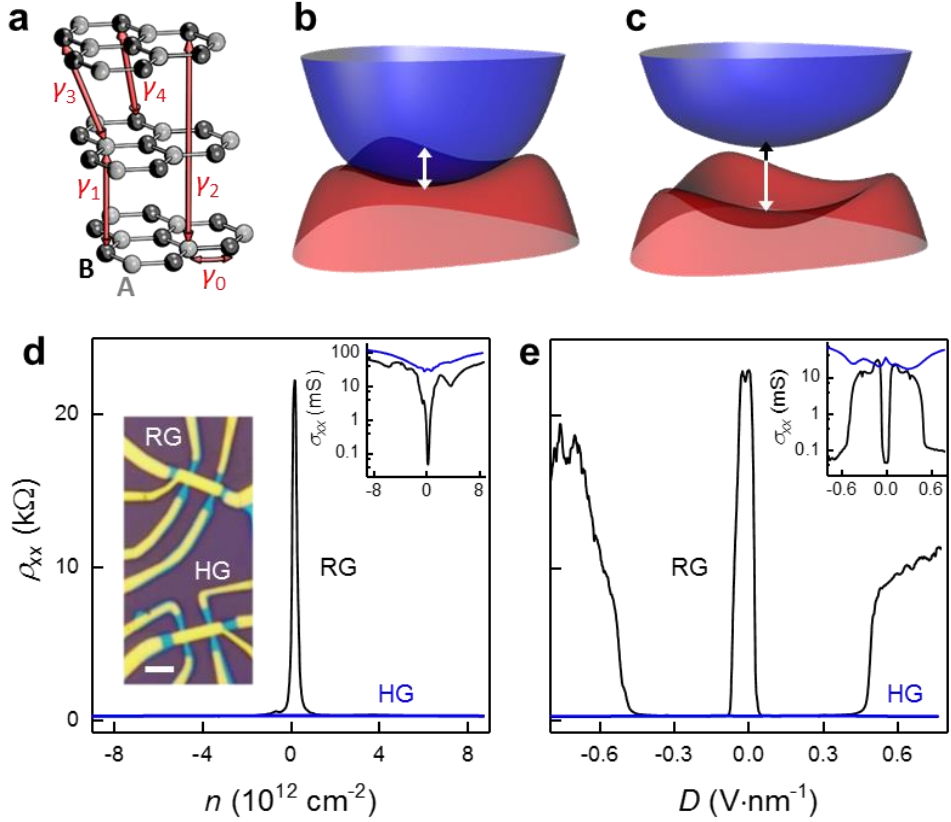

**Fig. 1: Transport characteristics of thin rhombohedral graphite film. a,** Schematic of RG graphite hopping parameters $\gamma_i$. **b,** Dispersion of the surface states including additional hopping parameters. The vertical axis is energy, $E$, and horizontal axes are in-plane momenta, $p$. Skew hopping $\gamma_4$ adds an appreciable bandwidth $2\gamma_4\gamma_1/\gamma_0$ to the otherwise flat surface state bands, and introduces electron-hole asymmetry. The presence of $\gamma_2$ and $\gamma_3$ leads to trigonal warping. **c,** Spontaneous band gap $\Delta$ in thin RG devices overcomes the band overlap, $\Delta > 2\gamma_4\gamma_1/\gamma_0$. **d,** $\rho_{xx}$ as a function of total carrier density $n = n_t + n_b$ measured at $D = 0$ for 3 nm thick crystals with rhombohedral (RG, black curve) and hexagonal (HG, blue curve) crystal structures. **e,** $\rho_{xx}$ as a function of $D$ measured at $n = 0$. Right insets in **d** and **e** are corresponding $\sigma_{xx}$ curves. Left inset in **d** shows the optical micrograph of the devices. Scale bar, 2 μm.

We identified the stacking order of graphite films using Raman spectroscopy and, to retain the high electronic quality, we encapsulated the exfoliated films with hexagonal boron nitride (hBN) crystals using directional dry transfer technique[7], see Supplementary Materials. Figure 1d compares low temperature resistivity $\rho_{xx}$ of RG and HG devices of the same thickness (≈ 3 nm thick, 9 graphene layers) as a function of total carrier density $n = \frac{C_t V_t + C_b V_b}{e}$, where $V_t$ and $V_b$ are the top and bottom gate voltages respectively, and $C_t$ and $C_b$ are the unit area capacitance of top and bottom gates respectively, $e$ the elementary charge. RG device (device 1) shows much stronger modulation of $\rho_{xx}(n)$ as compared to the HG device such that $\rho_{xx}$ of RG device 1 spans from 10 Ω to 23 kΩ and exhibits a clear high-resistivity region near $n = 0$. At the same time, the HG device displays metallic behaviour over the entire range of concentrations ($\rho_{xx}$ varies only from 8 Ω to 63 Ω), similarly to the reported dual-gated tetra- and hexa-layer ABA graphene[13,14]. Figure 1e shows three distinct high-resistance regions of RG device 1 (which are absent in the HG device) as a function of displacement field $D = \frac{C_t V_t - C_b V_b}{2\varepsilon_0}$ measured at $n = 0$ ($\varepsilon_0$ is the vacuum permittivity).

In high-resistivity regions of RG device 1, $\rho_{xx}$ is of the order of $h/e^2$ ($h$ is Planck's constant) suggesting that a band gap is opened in the surface states spectrum and the conductivity is edge limited. Resistivity peak at the neutrality point ($n = 0$, $D = 0$) was only observed in thin (< 4 nm) RG devices and cannot be explained within a free-particle model (Supplementary Materials). Before discussing this high-resistivity feature let us consider states induced by a large displacement field. These regions of high resistivity, observed in all RG devices, emerging above a critical displacement field, $|D| > D_c$, (see Fig. 2a-d) are in agreement with the theoretical prediction that a high enough displacement field opens a band gap in the surface states of RG[15]. The effect is similar to that observed in a bilayer graphene under the displacement field, but with much stronger nonlinear screening of charges on the two surfaces of RG (see Supplementary Materials).

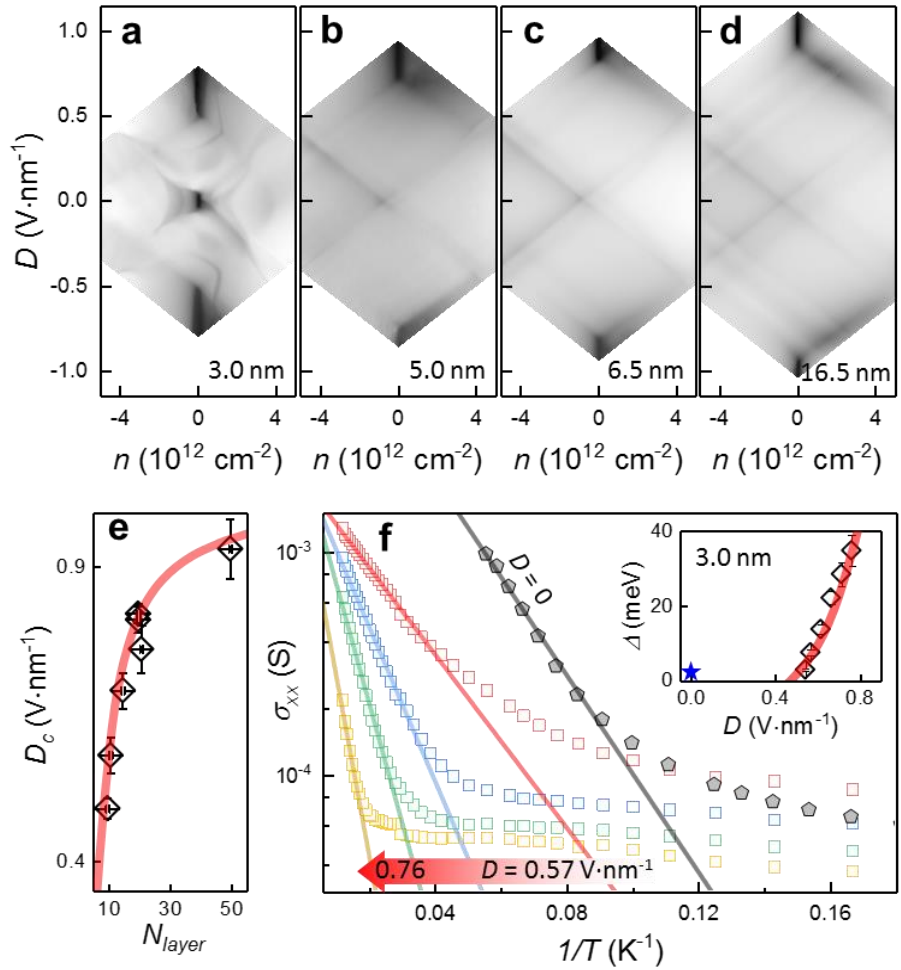

**Fig. 2: Thickness dependence of transport characteristics of rhombohedral graphite films. a-d,** Conductivity maps $\sigma_{xx}(n, D)$ for RG of different thicknesses. Logarithmic grayscales: black to white is 10 μS to 25 mS, 0.1 mS to 10 mS, 0.15 mS to 12 mS, and 0.4 mS to 100 mS for maps from **a** to **d**, respectively. **e,** Thickness dependence of the critical field $D_c$, above which the transport gap is opened (black diamonds). $D_c$ values were extracted from linear fits of $\sigma_{xx}(D)$ at $n = 0$ near the insulating states in the maps **a-d**. Red curve is the self-consistent model (see Supplementary Materials). **f,** Arrhenius plot of $\sigma_{xx}$ for the 3 nm RG at several $D$ from 0.57 V·nm$^{-1}$ (red squares) to 0.76 V·nm$^{-1}$ (yellow squares). Grey pentagons are the data for $D = 0$. Linear fits of high-temperature regions (solid lines) give the band gap $\Delta$ presented in the inset. Red curve in the inset is the band gap estimation using self-consistent screening model (see Supplementary Materials). For comparison, the thermal activation energy of the high-resistance state at $D = 0$ is also plotted in the inset (blue star).

The effect of RG thickness on the displacement-field-induced band gap is shown in Fig. 2a-d. It plots $\sigma_{xx}(n, D)$ maps for four RG devices with thicknesses ranging from 3 to 16.5 nm plotted (devices 1 to 4). Figure 2e summarises how the critical displacement field $D_c$ increases with the sample thickness. The solid line is the self-consistent theory (see Supplementary Materials). From the temperature dependence of $\sigma_{xx}$ of 3 nm thick RG device 1 we extracted the sizes of thermally activated gaps $\Delta$ at different $D$ (Fig. 2f), which are plotted in the inset in Fig. 2f. The band gap observed in RG at $|D| > D_c$ is in stark contrast to HG response to the displacement field where $D$ causes the resistivity to decrease (Fig. 1e) due to the growing size of the Fermi surfaces of the top and bottom surface states[15,16]. The behaviour of $\sigma_{xx}(D)$ presented in Figs. 2a-d is characteristic of the uniform ABCABC… stacking in our RG samples. Note that, if mixed ABC/ABA stacking was present in our devices, this would have prevented the gap opening by the displacement field, in agreement with theory (Supplementary Materials), which provided an independent confirmation of complete ABC stacking in our RG devices (Fig. S1 and S2).

Next we explore electron transport in RG films using magnetic field $B$ applied perpendicular to the device plane. Longitudinal $\rho_{xx}$ and transverse $\rho_{xy}$ resistivities of device 1 are shown in Fig. S3a as a function of $B$ at $n$ = 2.3·10$^{12}$ cm$^{-2}$ and $D$ = 0. The Shubnikov-de Haas (SdH) oscillations emerge at $B \approx 1$ T, and well-quantised $\rho_{xy}$ plateaux develop at $B > 3$ T manifesting the onset of quantum Hall effect (QHE), where under a strong magnetic field the electronic spectrum of 2D surface states collapses into a series of discrete Landau levels (LL), leading to the quantisation of $\rho_{xy}$. In high $B$ (see, e.g., Fig. S3b for $B$ = 10 T) $\rho_{xy}(n)$ plateaux appear at even fractions of $h/e^2$ indicating a two-fold degeneracy of the LLs. To examine the quantisation, we plot the Landau fan diagram $\sigma_{xx}(n, B)$ for device 1, Fig. 3a ($D$ = 0). Note a strong electron-hole asymmetry in the fan diagram: the electron side exhibits straight LLs (blue lines follow the slopes $B/n = h/ev$, where $v$ is the filling factor) fanning out from the neutrality point, while the hole side shows a criss-cross pattern. We observed a similar behaviour in the other RG devices of different thicknesses, which is consistent with the surface origin of the LLs.

In an $N$-layer thick RG film, a Berry phase $\pm N\pi$ around the opposite K points leads to $N$-fold degenerate zero energy Landau levels for each spin and valley, which are localised on the opposite surfaces in the opposite valleys[17,18]. A free-particle spectrum of Landau levels in RG calculated using the same method as in Ref[19] is plotted in Fig. S4. At low $B$, 0$^{th}$ LLs are degenerate with valence-band LLs, while at moderate $B$, conduction-band LLs curve up and valence-band LLs curve down, forming numerous crossings on the hole side of the spectrum, Fig. S4. In the absence of displacement field all LLs are valley degenerate. In addition, low-index valence-band LLs are triply-degenerate reflecting the triply-degenerate maxima of the valence band, and they are split at higher $B$ where the three valence band Fermi surfaces merge into a single one. For the 3 nm ($N$ = 9 layers) device in Fig. 3a, the total number of the 0$^{th}$ LLs is 36 (2 spins, 2 valleys, and 9 orbitals), and the lines corresponding to filling factors up to ±18 are coming from these 0$^{th}$ LLs. Lines on the electron side tracing $v$ = 8, 12, 16… are well pronounced and these filling factors in multiples of 4 can be explained by larger Zeeman gaps between two-fold degenerate orbital levels of 0$^{th}$ LLs. A pronounced gap at $v$ = -18 followed by a series of crossings below 8 T, could be traced back to the crossings between the 2$N$-fold degenerate 0$^{th}$ LL and the valence-band LLs. The robust $v$ = -18 state emerges when these crossings stop, Fig. S4. When only single gate is used, Fig. S5, the valley degeneracy is also lifted, and a robust gap appears at $v$ = -9 (for 9-layer-thick device 1), and at $v$ = -11 (for 11-layer-thick device 6), thus providing an additional support to the assignment of the spectrum to $N$-fold degenerate 0$^{th}$ LL as a manifestation of a giant Berry phase in RG films. Although the theory qualitatively agrees with the experiment, it predicts a twice higher value of $B$ at which the last crossings occur. The discrepancy is probably related to exchange interactions.

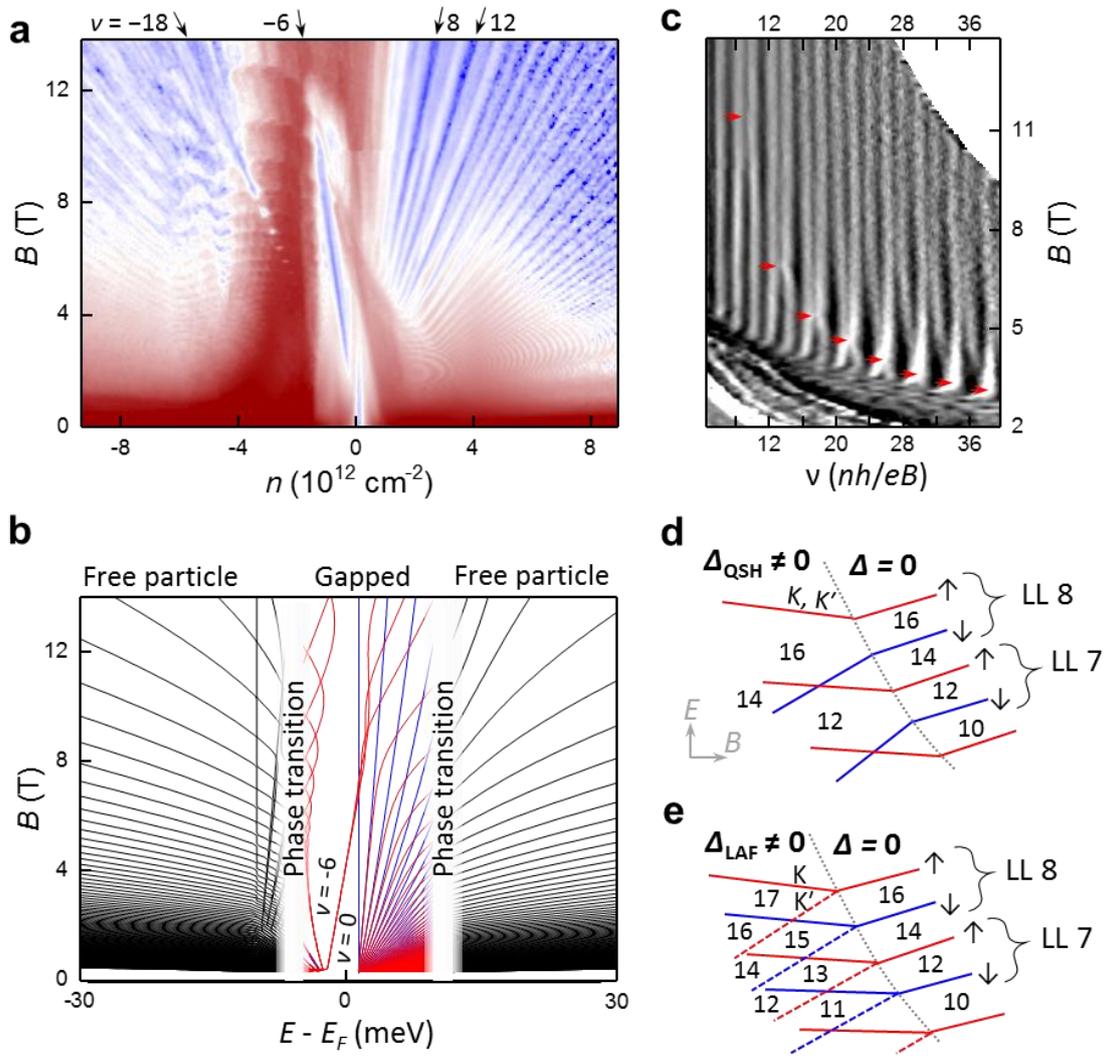

**Fig. 3: Quantum Hall effect in rhombohedral graphite. a,** Map of longitudinal conductivity $\sigma_{xx}$ as a function of $B$ and $n$ measured at $D = 0$ V·nm$^{-1}$, $T = 10$ mK. Logarithmic colour scale: blue to wine is 0.5 mS to 10 mS. **b,** Calculated spectrum of Landau levels in RG, using a gapped band structure at low energies, and the free particle theory at higher energies. Red and blue lines, depending on the order parameter, correspond to electronic states of opposite spins (quantum spin Hall), or of opposite valleys (layer antiferromagnetic). **c,** Differential $d\sigma_{xx}/d\nu(\nu, B)$ map for the electron side (the same fan diagram as in panel **a**). Red arrows mark the positions of Landau level crossings. Linear grey scale, black to white, -0.2 µS to 0.2 µS. **d,e,** Effect of quantum spin-Hall (QSH) (**d**) and layer antiferromagnetic (LAF) (**e**) order parameters on the Landau level spectrum. Numbers indicate the filling factors; LL 7 and 8 refer to orbital indices of the 0$^{th}$ Landau level. ↑ (red lines) and ↓ (blue lines) label spin-split LLs. Dotted lines demarcate phase transition between gapped and free particle regimes.

Although the free particle theory adequately describes the experimental results at high doping (cf. Fig. 3a and b, above $|n| > 2.5 \cdot 10^{12}$ cm$^{-2}$), it fails to explain the pronounced gaps observed at $\nu = 0$ and $\nu = -6$, Fig. 3a. Apart from the criss-cross pattern on the hole side, which agrees with the free-particle theory, there is a series of crossing points on the electron side at the symmetry broken states, $\nu = 10, 14, 18, 22…$ The crossings are better seen if we replot the Landau fan as a function of $\nu$ (Fig. 3c; the crossing points are marked with the red arrows). These crossings also cannot be explained using the free particle picture. To account for the latter features, we suggest a spontaneous band gap opening. Such a band gap – large enough to overcome the single-particle overlap between the conduction and valence bands – should split apart the conduction-band and valence-band LLs, thus allowing observation of robust quantum Hall states at $\nu = 0$ and -6 (Fig. 3b). The band gap disappears with increasing doping or magnetic field, but the observed crossings on the electron side (Fig. 3b,c) suggest that a small interaction-induced order parameter survives up to electronic doping $n=2.5 \cdot 10^{12}$ for 9-layer device.

The interplay between the orbital, valley, and spin degrees of freedom leads to a broad range of possible candidates for the spontaneous gap opening. RG is predicted to host spontaneous quantum Hall states, such as quantum valley Hall, quantum anomalous Hall, layer antiferromagnetic insulator, and quantum spin Hall states[3]. The crossing points on the electron side (Fig. 3c), observed only at our lowest attainable temperature (≈ 10 mK) can be explained by the phase transition from spin-polarised states due to Zeeman splitting at higher electron concentration to quantum spin Hall state at electron concentration $n < 2.5 \cdot 10^{12}$ cm$^{-2}$, Fig. 3d. Quantum spin Hall phase is accompanied by a giant Berry curvature and a related giant orbital magnetic moment (the same in both valleys but opposite for the two spin directions)[17]. The observed crossings cannot be explained by any other of the conjectured phases listed above, for instance, layered antiferromagnetic (LAF) phase leads to lifting the valley degeneracy and the presence of gaps and crossings at both even and odd filling factors, Fig. 3e, which contradicts the experiment.

In the absence of magnetic field, thin (< 4 nm) RG devices reveal an intrinsic insulating state at the neutrality point ($n$ = 0, $D$ = 0), Fig. 1d,e. The observed insulating state is unexpected in the single-particle picture[15,20,21]. When considering only $\gamma_0$ and $\gamma_1$ hopping amplitudes, surface states form flat bands, $E \approx \pm p^N$. However, the non-zero $\gamma_4$ leads to weakly dispersing and overlapping surface bands, Fig. 1b. The observed insulating state implies the presence of a band gap, $\Delta$, which exceeds the valence-conduction band overlap, $\Delta > 2\gamma_4\gamma_1/\gamma_0$, Fig. 1c. Figure S6a shows the temperature dependence of resistivity $\rho_{xx}(n)$ around the insulating state at $D$ = 0 in another thin (≈ 3.3 nm) RG film (device 5); its $\rho_{xx}(n_t, n_b)$ map is plotted in Fig. S6b. At temperatures below 10 K, the resistance shoots up by orders of magnitude, showing thermal activation behaviour consistent with a gap of 2-3 meV (inset in Fig. S6a). Note that this value represents an effective gap; the real bandgap includes the band overlap of the order of $2\gamma_4\gamma_1/\gamma_0$, Fig. 1c.

Surprisingly, we observed a strong hysteretic behaviour in the resistivity of thin RG devices around the neutrality-point insulating state, Fig. 4a (same device as in Fig. S6). The hysteresis rapidly decreases with the increase in temperature and disappears if the RG enters a metallic state. We also note that the hysteresis was observed only for the electron doping reflecting the strong electron-hole asymmetry of the band structure of RG films.

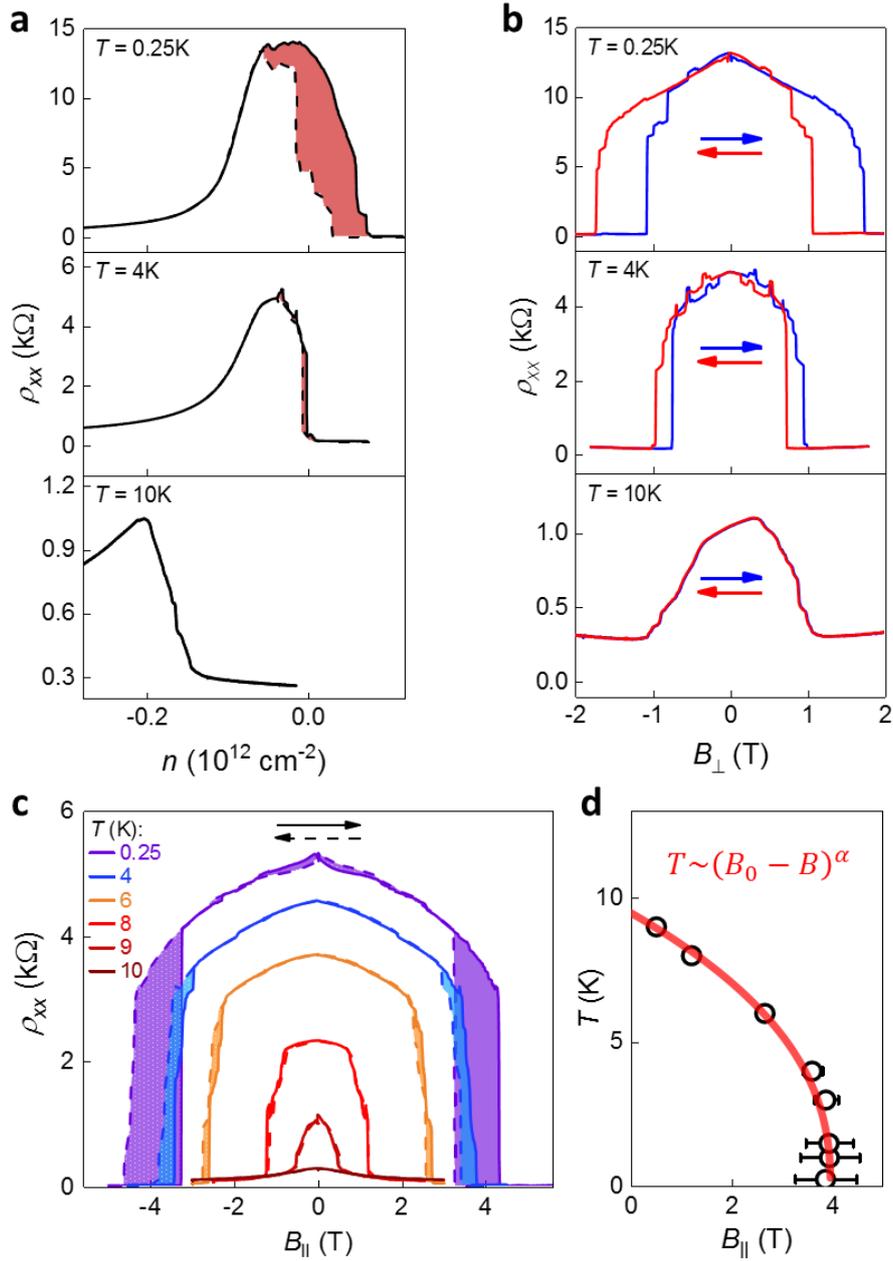

**Fig. 4: Temperature dependence of hysteretic behaviour of insulating state for thin RG devices. a,** $\rho_{xx}(n)$ for a dual-gated RG device of 3.3 nm thick (device 5) at $D = 0$ V·nm$^{-1}$. Solid (dashed) lines indicate positive (negative) sweep direction and the coloured areas highlight the hysteresis between the sweep directions. **b,** $\rho_{xx}$ hysteresis in perpendicular magnetic field for the same device 5; $n = 4 \cdot 10^{10}$ cm$^{-2}$ for $T = 0.25$ and 4.3 K, $n = -2.1 \cdot 10^{11}$ cm$^{-2}$ for $T = 10$ K. **c,** hysteresis in $\rho_{xx}$ for parallel magnetic field in a single-gated RG of 3.6 nm thickness (device 6). **d,** $B$-$T$ phase diagram of the critical behaviour of the order parameter based on the critical values of the parallel magnetic field below which the transition to insulating state occurs (shown in panel **c**) as a function of temperature (error bars represent the hysteresis). The red line in **d** is the fit $T \sim (B_0 - B)^\alpha$, where $B_0 \approx 4$ T is the critical field at $T = 0$ K, and $\alpha \approx 0.4$ is the critical exponent.

Besides the density dependence, we also observed strong hysteresis accompanied by sharp transitions, in response to both perpendicular ($B_\perp$) and parallel ($B_\parallel$) magnetic fields, Fig. 4b,c. $B_\perp \leq 2$ T suppresses the insulating state (Fig. 4b) but Landau quantisation complicates analysis of this temperature dependence. In contrast, measurements in the parallel field, Fig. 4c, where the orbital degree of freedom remains unaffected, provide a clear picture. With the increase in temperature, both the hysteresis and the transition field rapidly decrease, and the $B_\parallel$–$T$ phase diagram follows the power-law behaviour $T \sim (B_0 - B_\parallel)^\alpha$ (Fig. 4d), characteristic of strongly correlated electron systems[22].

The hysteretic behaviour, in combination with the insulating behaviour, hints towards a spontaneous formation of correlated electronic phases. The phases could be represented by mesoscopic scale domains, similar to the case of strongly correlated electronic systems in manganites, cuprates and transition metal oxides[23]. Such a hysteretic behaviour found in both electric and magnetic fields is a characteristic of multiferroic materials, and advanced scanning probe techniques could potentially help to visualise and analyse the electronic domains in the RG films.

In summary, RG films host robust topologically-protected surface states of high electronic quality. In the presence of a strong magnetic field, the quantum Hall effect behaviour shows a rich interplay between the spin, valley and orbital degrees of freedom. The presence of multiple interacting degrees of freedom (spin, charge, valley, and sublattice) is conducive to the electronic phase separation. Our work thus offers a novel experimental system to explore strongly-correlated phenomena, which does not rely on spin-orbit or exchange interactions of transition metal oxides[23], or careful structuring via moiré superlattices in twisted bilayer systems[24,25]. Electronic coupling between top and bottom surfaces in thin RG films, similar to that observed in thin films of topological insulators[26,27], could be responsible for the spontaneous gap opening in our systems, but this cannot explain the observed hysteretic behaviour. We believe that thin RG films provide an interesting playground to explore strong correlations, quantum criticality, and other many-body phenomena, usually reserved to *f*-elements-based heavy fermions or *d*-elements-based transition metal oxides.

## Acknowledgements
This work was supported by EU Graphene Flagship Program, European Research Council, the Royal Society, and Engineering and Physical Research Council (EPSRC). A.M. acknowledges the support of EPSRC Early Career Fellowship EP/N007131/1.

## References

1	Kopnin, N. B., Ijäs, M., Harju, A. & Heikkilä, T. T. High-temperature surface superconductivity in rhombohedral graphite. *Phys. Rev. B* **87**, doi:10.1103/PhysRevB.87.140503 (2013).
2	Otani, M., Koshino, M., Takagi, Y. & Okada, S. Intrinsic magnetic moment on (0001) surfaces of rhombohedral graphite. *Phys. Rev. B* **81**, doi:10.1103/PhysRevB.81.161403 (2010).
3	Zhang, F., Jung, J., Fiete, G. A., Niu, Q. & MacDonald, A. H. Spontaneous quantum Hall states in chirally stacked few-layer graphene systems. *Phys Rev Lett* **106**, 156801, doi:10.1103/PhysRevLett.106.156801 (2011).
4	Pamuk, B., Baima, J., Mauri, F. & Calandra, M. Magnetic gap opening in rhombohedral-stacked multilayer graphene from first principles. *Phys. Rev. B* **95**, 075422, doi:10.1103/PhysRevB.95.075422 (2017).
5	Muñoz, W. A., Covaci, L. & Peeters, F. M. Tight-binding description of intrinsic superconducting correlations in multilayer graphene. *Phys. Rev. B* **87**, doi:10.1103/PhysRevB.87.134509 (2013).
6	Xu, D. H. *et al.* Stacking order, interaction, and weak surface magnetism in layered graphene sheets. *Phys. Rev. B* **86**, doi:10.1103/PhysRevB.86.201404 (2012).
7	Yang, Y. *et al.* Stacking Order in Graphite Films Controlled by van der Waals Technology. *Nano Lett*, doi:10.1021/acs.nanolett.9b03014 (2019).
8	Ho, C. H., Chang, C. P. & Lin, M. F. Evolution and dimensional crossover from the bulk subbands in ABC-stacked graphene to a three-dimensional Dirac cone structure in rhombohedral graphite. *Phys. Rev. B* **93**, 075437, doi:10.1103/PhysRevB.93.075437 (2016).
9	Xiao, D., Chang, M. C. & Niu, Q. Berry phase effects on electronic properties. *Rev. Mod. Phys.* **82**, 1959-2007, doi:10.1103/RevModPhys.82.1959 (2010).
10	Lee, Y. *et al.* Competition between spontaneous symmetry breaking and single-particle gaps in trilayer graphene. *Nat Commun* **5**, 5656, doi:10.1038/ncomms6656 (2014).

11   Myhro, K. *et al.* Large tunable intrinsic gap in rhombohedral-stacked tetralayer graphene at half filling. *2D Materials* **5**, doi:10.1088/2053-1583/aad2f2 (2018).
12   Chen, G. *et al.* Tunable Correlated Chern Insulator and Ferromagnetism in Trilayer Graphene/Boron Nitride Moire Superlattice. *arXiv* (2019).
13   Shi, Y. *et al.* Tunable Lifshitz Transitions and Multiband Transport in Tetralayer Graphene. *Phys Rev Lett* **120**, 096802, doi:10.1103/PhysRevLett.120.096802 (2018).
14   Nakasuga, T. *et al.* Low-energy band structure in Bernal stacked six-layer graphene: Landau fan diagram and resistance ridge. *Phys. Rev. B* **99**, doi:10.1103/PhysRevB.99.085404 (2019).
15   Koshino, M. Interlayer screening effect in graphene multilayers with ABA and ABC stacking. *Phys. Rev. B* **81**, 125304, doi:10.1103/PhysRevB.81.125304 (2010).
16   Yin, J. *et al.* Dimensional reduction, quantum Hall effect and layer parity in graphite films. *Nature Phys.* **15**, 437-+, doi:10.1038/s41567-019-0427-6 (2019).
17   Koshino, M. & McCann, E. Trigonal warping and Berry's phase Nπ in ABC-stacked multilayer graphene. *Phys. Rev. B* **80**, doi:10.1103/PhysRevB.80.165409 (2009).
18   Min, H. K. & MacDonald, A. H. Chiral decomposition in the electronic structure of graphene multilayers. *Phys. Rev. B* **77**, 155416, doi:10.1103/PhysRevB.77.155416 (2008).
19   Slizovskiy, S., McCann, E., Koshino, M. & Fal'ko, V. I. Films of rhombohedral graphite as two-dimensional topological semimetals. *arxiv.1905.13094* (2019).
20   Xiao, R. J. *et al.* Density functional investigation of rhombohedral stacks of graphene: Topological surface states, nonlinear dielectric response, and bulk limit. *Phys. Rev. B* **84**, 165404, doi:10.1103/PhysRevB.84.165404 (2011).
21   Mcclure, J. W. Electron Energy Band Structure and Electronic Properties of Rhombohedral Graphite. *Carbon* **7**, 425-+, doi:Doi 10.1016/0008-6223(69)90073-6 (1969).
22   Ran, S. *et al.* Phase diagram of URu2-x Fe x Si2 in high magnetic fields. *Proc Natl Acad Sci U S A* **114**, 9826-9831, doi:10.1073/pnas.1710192114 (2017).
23   Dagotto, E. Complexity in strongly correlated electronic systems. *Science* **309**, 257-262, doi:10.1126/science.1107559 (2005).
24   Cao, Y. *et al.* Correlated insulator behaviour at half-filling in magic-angle graphene superlattices. *Nature* **556**, 80-84, doi:10.1038/nature26154 (2018).
25   Cao, Y. *et al.* Unconventional superconductivity in magic-angle graphene superlattices. *Nature* **556**, 43-50, doi:10.1038/nature26160 (2018).
26   Zhang, Y. *et al.* Crossover of the three-dimensional topological insulator Bi2Se3 to the two-dimensional limit. *Nature Phys.* **6**, 584-588, doi:10.1038/nphys1689 (2010).
27   Neupane, M. *et al.* Observation of quantum-tunnelling-modulated spin texture in ultrathin topological insulator Bi2Se3 films. *Nat Commun* **5**, 3841, doi:10.1038/ncomms4841 (2014).
28   Cai, S. *et al.* Independence of topological surface state and bulk conductance in three-dimensional topological insulators. *Npj Quantum Mater* **3**, 62, doi:10.1038/s41535-018-0134-z (2018).
29   Dresselhaus, M. S. & Dresselhaus, G. Intercalation compounds of graphite. *Adv. Phys.* **51**, 1-186, doi:10.1080/00018730110113644 (2002).
30   Garcia-Ruiz, A., Slizovskiy, S., Mucha-Kruczynski, M. & Fal'ko, V. I. Spectroscopic Signatures of Electronic Excitations in Raman Scattering in Thin Films of Rhombohedral Graphite. *Nano Lett* **19**, 6152-6156, doi:10.1021/acs.nanolett.9b02196 (2019).

# Supplementary Materials

## S1. Encapsulated rhombohedral graphite films and device fabrication

The RG films were obtained by mechanical exfoliation of natural graphite flakes (NGS naturgraphit, Graphenium Flakes 25-30mm) onto $SiO_2$ (290nm) / Si wafer. Exfoliated graphite films often show domains of various stacking orders, which can be identified by Raman spectroscopy. By mapping the ratio of the integral area of the low frequency component (ranging around 2670-2700 cm$^{-1}$) and high frequency component (ranging around 2700-2730 cm$^{-1}$) of Raman 2D band, we can visualise the distribution of rhombohedral stacking of the films[8]. The Raman spatial maps were taken with a step size of 0.5 μm.

The RG films often revert back into Bernal stacking during the flake transfer process, and only a small region of the flake remains in the original stacking order. The transition between rhombohedral stacking and Bernal stacking occurs during the shift of graphene layers along armchair directions, whereas it does not happen along zigzag directions. To retain high electronic quality of RG, we used directional van der Waals assembly technique[8] as follows. Firstly, we identified the crystal orientation of the graphite films from the edge chirality of graphite films using Raman spectroscopy. Next, we used polymer (polymethylmethacrylate (PMMA) / polydimethylsiloxane (PDMS) stack mounted on a glass slide) mediated dry transfer method to encapsulate RG films by hexagonal boron nitride (hBN). During the encapsulation process, the pressing down direction of the PDMS/PMMA layers was controlled along the zigzag direction of graphite films, so that the stacking transformation could be avoided. Then, we used Raman ratio map again to confirm the rhombohedral stacking of the graphite film encapsulated by hBN. Finally, the dual-gate multi-terminal Hall bar devices were made onto the hBN/RG films/hBN heterostructures using standard electron-beam lithography process.

## S2. Surface and bulk contributions to transport properties

Fig. S7 shows the temperature dependence of the resistivity $\rho_{xx}(T)$ at $n = 0$ of 16.5 and 3.3 nm thick RG samples (devices 4 and 5, respectively). In contrast to the metallic behaviour of Bernal-stacked graphite[16], we observed a semiconductor-to-metal-transition with decreasing the temperature. We attribute this transition to the interplay between three parallel conduction channels: the bulk, and the two surfaces. Infinitely thick RG has 3D Dirac cones in its bulk energy spectrum[8], but in films of a finite thickness ($N$ layers), these Dirac cones are gapped out by ≈ 2.1 eV/$N$. At high temperatures, the bulk of RG conducts as a result of thermal activation of 3D bands across the gap and dominates over the 2D surface states. With decreasing temperature, the contribution from the bulk decreases and, finally, below a critical temperature $T^*$ (≈ 20 K and 40 K for 16.5 nm and 3.3 nm RG, respectively), the contribution from the surfaces dominates as $\rho_{xx}(T)$ shows metallic behaviour. This temperature behaviour of resistivity is similar to that observed in 3D topological insulators, which is also ascribed to competing conductances of bulk and surface states[28]. The observed $T^*$ is smaller for thicker samples in agreement with 1/$N$ dependence of the bulk band gap. The observed temperature dependence indicates that at low temperatures ($T < T^*$) the conductivity through 3D bulk bands is suppressed and only 2D surface states contribute to the electronic transport.

## S3. Tight-binding model for RG films and electronic dispersion of surface states

To describe RG film we employ the Slonczewski-Weiss-McClure (SWMC) parametrization of graphite[17,21,29,19], with interlayer hopping parameters that take into account the lattice arrangement for rhombohedral graphite, as displayed in Fig. 1a. RG films have bulk subbands shifted away from zero energy, leading to a bulk gap of ≈ $3\pi\gamma_1/N$ for $N$ layers, making only the surface bands to be relevant for electronic transport. The surface states have shallow dispersion for quasi-momenta $p < p_c = \gamma_1/v$, where $v$ is Dirac velocity determined by intralayer nearest-neighbour hopping $\gamma_0$. There is also a quadratic term in the dispersion of surface states[19], $p^2/(2\,m_*)$ with $m_*^{-1} = 2\,v^2/\gamma_1(2\,\gamma_4/\gamma_0 + \delta/\gamma_1) \approx (0.4\,m_e)^{-1}$, that breaks the electron-hole symmetry, and trigonal warping terms controlled by $\gamma_2$ and $\gamma_3$. Since the SWMC parameters are poorly known for RG, we use the conventional HG values[16,29] for the larger hopping amplitudes $\gamma_0$ = 3.16 eV, $\gamma_1$ = 0.39 eV, $\gamma_3$ = 0.25 eV while tuning the unknown $\gamma_2$ = -0.01 eV and $m_* = 0.4\, m_e$ (we set the energy difference of bulk and surface layers to zero, $\delta = 0$, and tuned $\gamma_4 \approx 0.13$ eV) to better describe the experiment.

## S4. Band gap induced by displacement field

Introducing a vertical electric displacement field *D* via asymmetric gating leads to a potential difference $\tilde{\Delta}$ between the top and the bottom layers, resulting in a density redistribution that strongly screens the external electric field. The dispersion of surface states with non-zero $\tilde{\Delta}$ is shown in Fig. S8. Note that the difference between the energies of valence and conduction band states at the *K* point equals to $\tilde{\Delta}$ (only $\gamma_1$ vertical hopping and $\tilde{\Delta}$ potential contribute to the Hamiltonian at *K* points, leading to states localised at $A_1$ and $B_N$ sublattices with energies $\pm\tilde{\Delta}/2$, while all the other states belong to dimers, connected by $\gamma_1$ vertical hopping and having energies $\pm\gamma_1$ ). After calculating the energy dispersion, the screening electron density accumulated at layer *i* is calculated as

$$n_i = 2 \int_{BZ} \frac{d^2 k}{(2\pi)^2} \sum_{l=1}^{2N} \left[ \left( \left|\Psi_{A_i}^l(k)\right|^2 + \left|\Psi_{B_i}^l(k)\right|^2 \right) f(\varepsilon_l - E_F) - 1 \right]$$

where *i* is the layer number in *N*-layer RG, *l* stands for band index, $\Psi_{A_i}^l(k)$ and $\Psi_{B_i}^l(k)$ are normalized wave functions of two sublattices in layer *i*, *f* is the Fermi function, $E_F$ is Fermi level, and -1 subtracts a homogeneous charge jelly that neutralizes the Fermi electron density. Denoting the total screening charge, accumulated near the top and bottom surfaces by $n_t^{\text{scr}}(\tilde{\Delta})$ and $n_b^{\text{scr}}(\tilde{\Delta})$ respectively, we relate the displacement field to the charge density and the potential asymmetry as

$$\frac{-e[n_t - n_b]}{2\, \varepsilon_0} \equiv D = \frac{e\left[n_t^{\text{scr}}(\tilde{\Delta}) - n_b^{\text{scr}}(\tilde{\Delta})\right]}{2\, \varepsilon_0} + \frac{\varepsilon\tilde{\Delta}}{d}$$

where *e* is the electron charge, *d* is the thickness of graphite film, and $\varepsilon_0$ is vacuum permittivity. We employed Hartree approximation and used the fact that the dominant screening charge is localised at the surface layers. By varying the potential difference of surface layers, tuning the chemical potential to maintain overall charge neutrality, and finding the electron density distribution we determined the relation between the band gap and applied external displacement field, as shown in the inset of Fig. 2f. The band gap $\Delta$, defined as the difference between the conduction band minimum and the valence band maximum, is shown in Fig. S8.

In the absence of displacement field *D*, the conduction and valence bands of RG overlap. Therefore, one has to apply a field *D* larger than some critical $D_{crit}$, so that the potential asymmetry exceeds the bandwidth of overlapping and drives the system from semimetallic to semiconducting phase, Fig. S8. Calculated values of $D_{crit}$ are shown in Fig. 2e and compared to experimental results. When the total electron concentration is zero, $n_t - n_b = 0$, the Fermi level is in the gap, resulting in the observed experimentally transport gap presented in the main text.

## S5. Possibility of stacking faults

It is important to have a reliable method to check for the possible stacking faults in the sample. Apart from Raman[7] (Fig. S1a,b) and electronic Raman scattering signatures[30], measurement of an in-plane transport gap opened by the applied vertical displacement field (asymmetric gating) provides another diagnostic tool (Fig. S1c-e). Our calculations reveal that it is impossible to open a transport gap if stacking faults or thicker inclusions of ABA stacking are present. We considered the most elementary "twin-boundary" type of stacking fault that joins ABC stacking with its mirror-image CBA stacking as illustrated in Fig. S2a for A**BCB**A penta-layer, a Bernal stacking fault that joins two ABC multilayers, as shown is Fig. S2d for A**BCBC**A multilayer, and a surface stacking fault AB**CAC**, Fig. S2g.

At K points of Brillouin zone, the in-plane hoppings $\gamma_0$ destructively interfere, leaving only $\gamma_1$ as high-energy scale. For the twin-boundary stacking, the low-energy states are localised on sublattices $A_1$, $A_5$, and $B_3$, and on a zero-energy eigenstate $(\Psi_{B2}, \Psi_{B4}) = \left(\frac{1}{\sqrt{2}}, -\frac{1}{\sqrt{2}}\right)$ of a $B_2A_3B_4$ trimer, which we denote as $B_2$-$B_4$, Fig. S2a. Away from the K point, the chiral decomposition[18] shows an appearance of two effective 2- and 3-layer ABC chiral subsystems with obvious generalisation to thicker films. So, there are four low-energy sub-bands, two with energies growing away from K point (conduction-band-type) and two with energies decreasing away from K (valence-band-type). Similar to RG discussed before, the hopping $\gamma_4$ leads to an overlap of the order of 20 meV of valence and conduction sub-bands. Application of the vertical electric field (creating layer potentials $U_1 < U_2 < ... < U_5$) splits the four almost degenerate sub-bands into two outer valence and conduction bands, localised on $A_1$ and $A_5$ (with energies $U_1$ and $U_5$ at K point), and two inner valence and conduction bands, localised on $B_3$ and $B_2$-$B_4$ (with energies $U_3$ and $(U_2 + U_4)/2$ at K point). The energy splitting of the two inner bands is $U_3 - (U_2 + U_4)/2 = \frac{e^2 d_1 n_{B3}}{2 \epsilon_0}$, $d_1$ = 0.335 nm, and it only depends on the charge density of the third layer.

For the Bernal A**BCBC**A stacking fault, Fig. S2d, the low-energy states at the K point are localised on sublattices $A_1$, $B_3$, $A_4$, and $B_6$, with the splitting of inner band determined by the electric field between layers 3 and 4. In both stacking fault cases, to open a gap one requires a displacement field sufficient to overcome the doubled screening of four low-energy bands and, additionally, to create ≈ 20 meV potential difference between the two consecutive layers. Such displacement fields would strongly exceed the values plotted in **Fig. 2e,** and are, typically, beyond experimental reach. Large but experimentally attainable displacement fields would split the inner $B_3$, $A_4$ bands at the K point, as shown in Fig. S2e,f, but such high fields would also significantly increase the dispersion of these bands, resulting in the band overlap near $p = p_c$, thus closing the gap. As for the ABA surface stacking fault, Fig. S2g-I, the ABA surface bands overlap with a Dirac cone having velocity $v/\sqrt{2}$, which shunts the resistive state.

Thus, we conclude that any stacking faults make opening of a transport gap impossible. We also note that stacking faults would, generically, lead to asymmetric behaviour with respect to reversal of the direction of *D.* Calculated dispersions with non-zero displacement field are shown for the twin-boundary defect in Fig. S2b,c, for the stacking fault buried inside the ABC bulk in Fig. S2e,f, and for the surface stacking fault in Fig. S2h,i. Experimental data of a sample where the stacking fault is present are displayed in Fig. S1.

### S6. Calculation of the spectrum of Landau levels

We determine the spectrum of Landau levels (LLs) employing the same method as used in Ref[19]. Numerical diagonalization is performed in a regularised finite basis consisting of oscillator states for each sublattice component, $(\phi_0, \phi_1, ..., \phi_{N_0+n-1})$ for sublattice $A_n$ and for $(\phi_0, \phi_1, ..., \phi_{N_0+n})$ for $B_n$, where $N_0$ = 250 is a cutoff Landau level index, sufficient for convergence at $B$ > 0.5 T.

To qualitatively understand the LL spectrum, let us start with the minimal model[17] with only the nearest-neighbour in-plane ($\gamma_0$) and interlayer ($\gamma_1$) hoppings. In the absence of displacement field ($D$ = 0), the low-energy LL spectrum consists of 4$N$ degenerate levels at zero energy, which we call by $0^{th}$ LLs (where 4 is due to double-spin and double-valley degeneracy, and $N$ comes from $N$-fold orbital degeneracy), plus four-fold degenerate levels arising from the conduction and valence bands that disperse as $E^{\pm} \sim \pm B^{N/2}$. Within the two-band model, the vector wavefunctions of the $0^{th}$ LLs of one valley reside on one (e.g., top) surface, $\Psi_{K^+,i} = \begin{pmatrix} \varphi_i \\ 0 \end{pmatrix}, i = 0, ..., N-1$, while for the other valley the wavefunctions sit on the opposite (e.g.

bottom) surface, $\Psi_{K^-,i} = \begin{pmatrix} 0 \\ \varphi_i \end{pmatrix}$. Therefore, for 0$^{th}$ LLs the valley degeneracy is equivalent to the top/bottom surface degeneracy.

The minimal model is insufficient to describe the experimental results, making it necessary to consider further electron hopping amplitudes. The presence of inter-layer same-sublattice hopping $\gamma_4$ leads to strong electron-hole asymmetry and breaks the orbital degeneracy of 0$^{th}$ LLs, leading to dispersion $E_n^0 = n\hbar\omega_c, n = 0, \ldots, N-1$, with $\omega_c = eB/m_*$, $m_* \approx 0.4\,m_e$. The conduction band LLs continue the sequence of fourfold degenerate LLs as $E_n^+ = n\hbar\omega_c, n = N, N+1, \ldots$. The valence band dispersion is non-monotonic and produces a series of fourfold degenerate LLs $E_n^- = n\hbar\omega_c, n = 1,2,\ldots$ near the zero energy, as well as a sequence of 12-fold degenerate levels at higher energy originating from the three equivalent valence band maxima, caused by trigonal warping (hoppings $\gamma_3$, $\gamma_2$).

At moderate magnetic fields, valence band LLs bend down, forming numerous crossings with 0$^{th}$ and conduction band LLs[17]. This explains multiple crossings of the LLs on the hole side in Fig. 3a. The twofold degeneracy of the filling factors (e.g. $v$ = 6, 8, 10…) at *D*=0 can be attributed to preserved valley degeneracy, while the gaps correspond to orbital splitting ($v$ = 6, 10, 14… for odd *N* and $v$ = 4, 8, 12, 16… for even *N*) and spin splitting ($v$ = 4, 8, 12, 16… for odd *N* and $v$ = 6, 10, 14… for even *N* ).

The filling factor at a specified chemical potential is determined as a number of filled LLs in the regularised basis, with half of the total number of levels subtracted, to account for background charge neutralising the Fermi sea. Normally, this would place the zero electron density between LLs originating from valence and conduction bands that form pairs, related by approximate electron-hole symmetry $(\Psi_A, \Psi_B) \to (\Psi_A, -\Psi_B)$, but, similarly to monolayer graphene, there are *N* additional (spin and valley degenerate) zero LLs (numbered as *0,…,N-1*) with wave-function localised only on one of the sublattices. At low magnetic fields, zero LLs overlap with valence-band levels, leading to complicated counting of filling factor, but at magnetic fields $B \gtrsim 10$ T, conventional valence band LLs are below the zero energy, leading to zero electron density ($v = 0$) located in the middle point of 0$^{th}$ LLs. This middle point is located either in between the Zeeman-split 0$^{th}$ LLs with index $(N-1)/2$ for odd *N*, or between the orbital-split 0$^{th}$ LLs with indices $N/2 - 1$ and $N/2$ for even *N*.

### S7. Spin- and valley-polarised states in the quantum Hall regime

In this section we examine the implications of spontaneous onset of quantum layer antiferromagnetic (LAF) and quantum spin Hall (QSH) orders on the spectrum of Landau levels in quantum Hall regime (we exclude the anomalous Hall order out of consideration as the anomalous Hall signal has not been observed). The possible mean-field order parameters are given by[3] $\Delta^{LAF} = \lambda\,\sigma_z^{layer} \otimes \sigma_z^{spin}$ and $\Delta^{QSH} = \lambda\,\sigma_z^{layer} \otimes \sigma_z^{spin} \otimes \sigma_z^{valley}$ where $\sigma_z$ is a Pauli matrix. Both the displacement field and the above order parameters lead to giant Berry curvature and a related giant orbital magnetic moment, peaked near the three degenerate maxima of the valence band[17]. The orbital magnetic moment has opposite sign in the two valleys for $\Delta^{LAF}$ or for gate-induced displacement field, while the orbital magnetic moment becomes the same in both valleys (but opposite for the two spin directions) for $\Delta^{QSH}$. The orbital magnetic moment leads to strong lifting of 12-fold degeneracy of LLs, originating from valence band maxima, leading to two groups of 6-fold degenerate LLs, explaining a robust gap at $v = -6$, Fig. 3a,b. The 6 filled valence band LLs correspond to $K \uparrow$ and $K' \downarrow$ in case of LAF order parameter, or to a ferromagnetic state $K \uparrow$ and $K' \uparrow$ in case of QSH order parameter.

The effect of order parameters on the spectrum of 0$^{th}$ LLs amounts to valley-/spin- dependent shift for LAF/QSH phases, which follows from localization of 0$^{th}$ LLs on the top/bottom surface in K/K' valleys. Hence, $\Delta^{LAF}$ breaks the valley degeneracy, leading to appearance of gaps at odd filling factors, Fig. 3e, while the valley degeneracy remains intact for $\Delta^{QSH}$, Fig. 3d. Experimental data in Fig. 3c are consistent with the onset of QSH order for electron doping below $2.5 \cdot 10^{12}$ cm$^{-2}$, where the orbital and spin magnetic moments align in parallel.

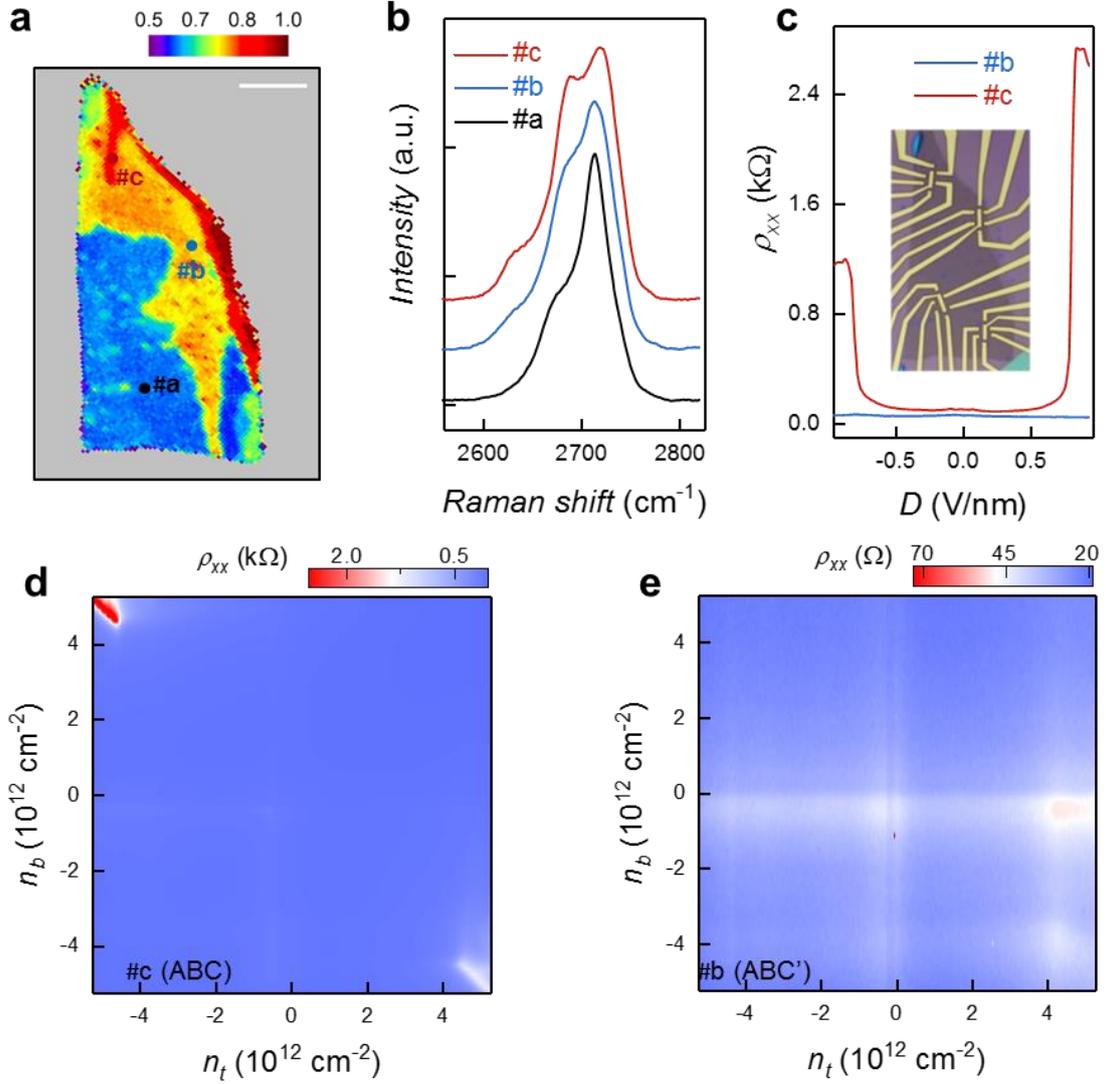

**Fig. S1: Effect of stacking sequence on the displacement field induced band gap. a,** Raman map of a 6.5 nm graphite flake with different domains regarding to their local stacking sequence. The colour coding follows the ratio of the integral area of the low frequency component (ranging around 2670-2700 cm$^{-1}$) to the high frequency component (ranging around 2700-2730 cm$^{-1}$) of graphite Raman 2D band. The coloured dots mark the positions where the Raman spectra shown in **b** were taken. Scale bar is 10 μm. **b,** Typical 2D Raman peak of HG (#a, black curve), RG (#c, red curve), as well as graphite of mixed ABA and ABC stacking (#b, blue curve). **c,** $\rho_{xx}$ as a function of $D$ for Hall bar devices made in domains the same as #b (light blue) and #c (red), respectively, $T$ = 1.6 K. Inset shows the optical micrograph of the devices before shaping them to Hall bar geometry. **d,e,** Resistivity maps $\rho_{xx}(n_t, n_b)$ of graphite with ABC stacking (d) and mixed stacking (e), respectively, $T$ = 1.6 K, $B$ = 0 T.

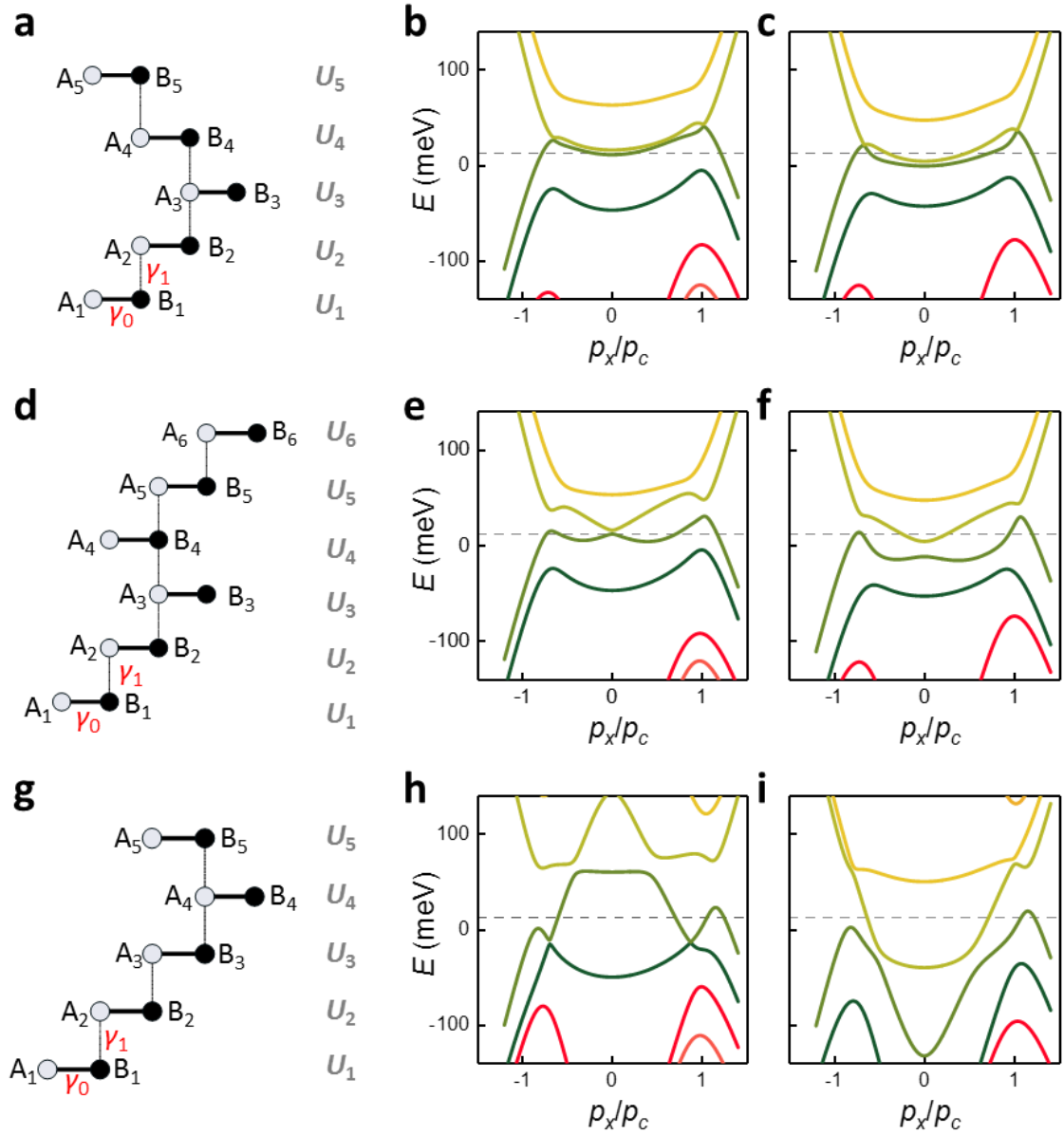

**Fig. S2: Calculated band structures of mixed-stacking graphite films (≈ 6.5 nm thick) in the presence of a large displacement field (|*D*| > ±0.9 V·nm$^{-1}$) show the absence of a *D*-field induced band gap. a,** Schematic of twin-boundary type defect ABCBA, **b,c,** Calculated band structures of twinned ABCABC**ABA**CBACBACBA sequence at positive (**b**) and negative (**c**) *D*, **d,** Schematic of a buried Bernal stacking fault ABCBCA, **e,f,** Calculated band structures of buried defect ABCABC**ABAB**CABCABCA sequence at positive (**e**) and negative (**f**) *D*, **g,** Schematic of a surface stacking fault ABCAC, **h,i** – bands of ABCABCABCABCABC**ABA** sequence at positive (**h**) and negative (**i**) *D*. Dashed line marks the position of the Fermi level at the charge neutrality.

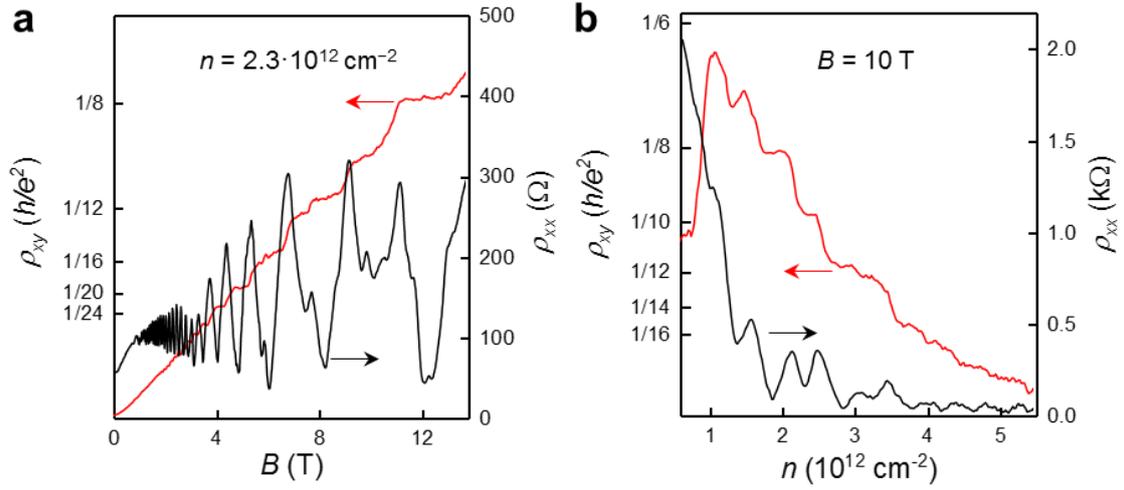

**Fig. S3: Quantum Hall effect in rhombohedral graphite. a**, Hall resistivity $\rho_{xy}$ (red curve) and longitudinal resistivity $\rho_{xx}$ (black) as a function of magnetic field $B$ measured at 10 mK in the same RG device presented in Fig. 1a in the main text (device 1), $n$ = 2.3 x $10^{12}$ cm$^{-2}$, $D$ = 0 V·nm$^{-1}$. **b**, $\rho_{xy}$ and $\rho_{xx}$ as a function of $n$, $D$ = 0 V·nm$^{-1}$, $B$ = 10 T, $T$ = 10 mK.

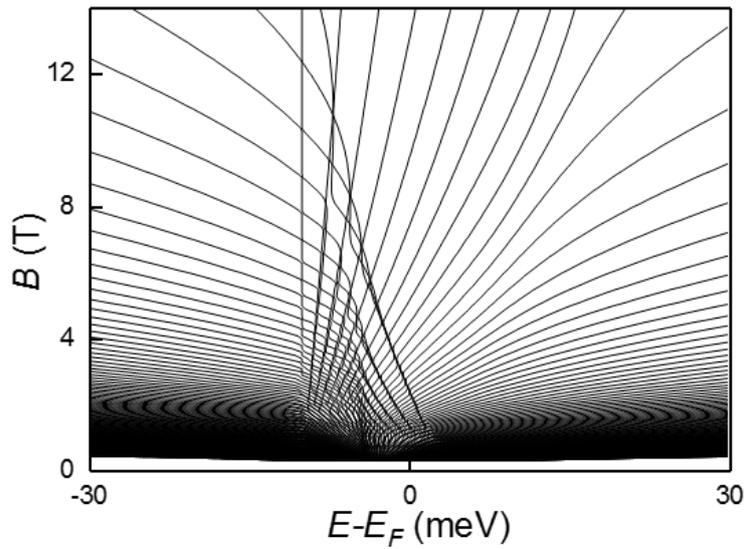

**Fig. S4: Free-particle spectrum of Landau levels in RG films.**

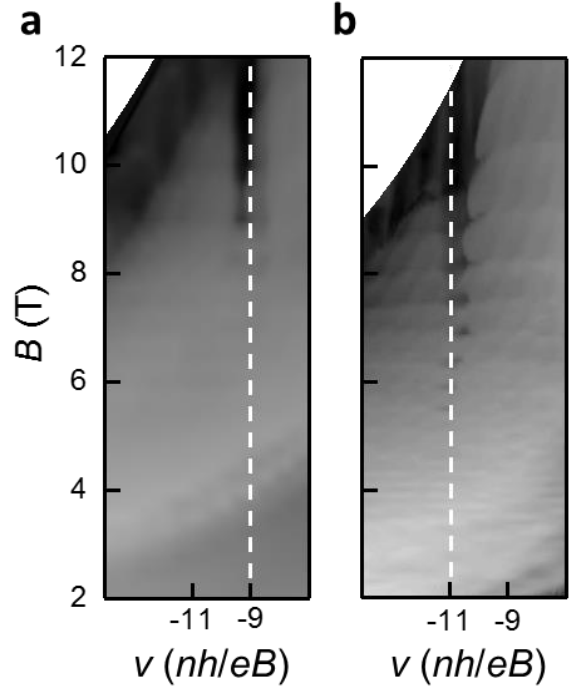

**Fig. S5: Single-gate ($D \neq 0$) Landau fan diagrams highlighting the robust ν = -$N$ quantum Hall state in $N$-layer-thick RG films. a,** Conductivity map $\sigma_{xx}(\nu, B)$ for 9-layer-thick RG film (device 1). **b,** $\sigma_{xx}(\nu, B)$ for 11-layer-thick RG film (device 6). Logarithmic grayscales from 10 μS to 10 mS, and 5 μS to 10 mS, for panels (**a**) and (**b**) respectively.

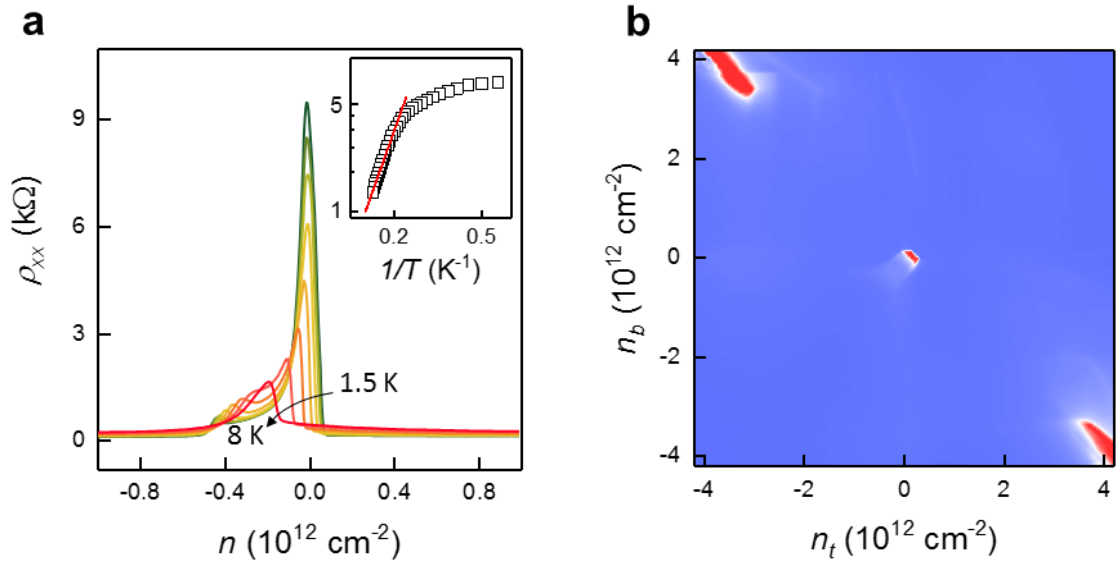

**Fig. S6: Observation of the insulating state in 3.3 nm RG film (device 5), same device as in Fig. 4a,b in the main text. a,** Temperature dependence of $\rho_{xx}(n)$ around the insulating state, $B = 0$ T. Inset is the Arrhenius plot of the peak resistivity, indicating the presence of a band gap ≈ 2-3 meV. **b,** Low-temperature (0.3 K) resistivity map $\rho_{xx}(n_t, n_b)$. $n_t$ and $n_b$ are carrier density induced by applying a top-gate voltage and back-gate voltage, respectively. Linear colour scale; light blue to red is 10 Ω to 6 kΩ.

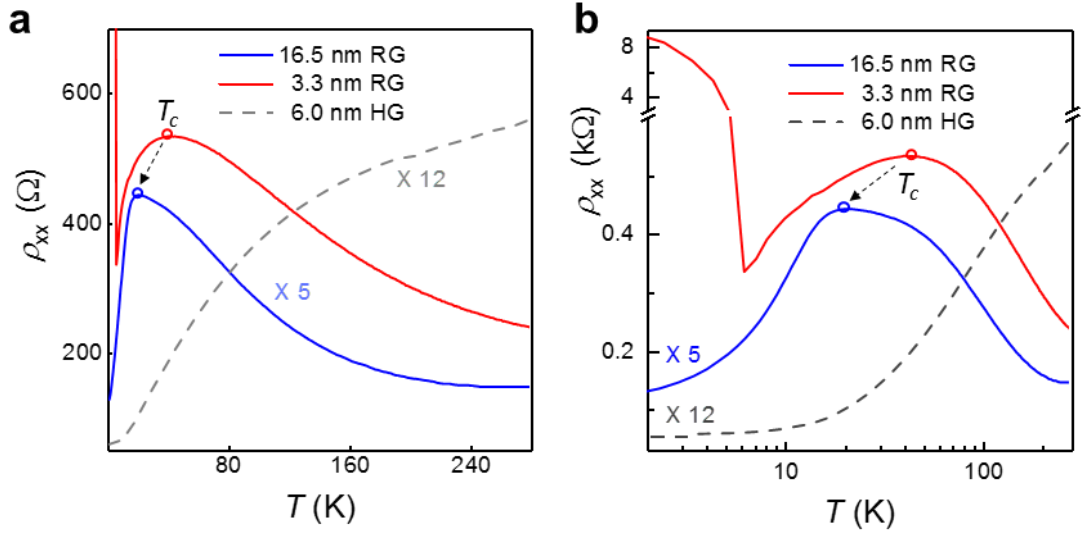

**Fig. S7: Temperature dependence of resistivity.** $\rho_{xx}$ as a function of $T$ at zero gate doping for RG with a thickness of 16.5 nm (blue solid curve) and 3.3 nm (red solid curve), as well as a 6 nm HG (grey dashed line). While cooling down, $\rho_{xx}$ of RG first increases for $T > T_c$, and then decreases, in sharp contrast to the monotonous decrease of $\rho_{xx}$ for HG. The critical temperature $T_c$ decreases with increasing thickness of RG. Besides the presence of $T_c$, $\rho_{xx}$ of the 3.3 nm RG shows a sharp increase for $T < 6$ K, due to phase transition to the insulating state.

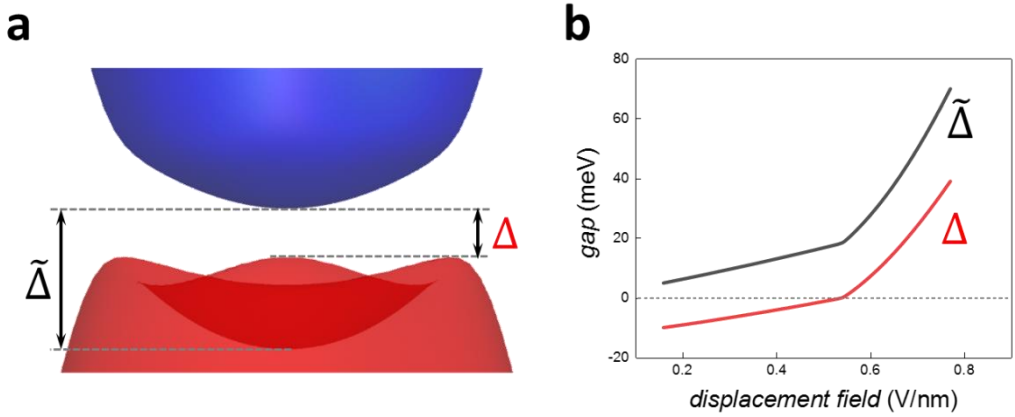

**Fig. S8: Band gap opening by displacement field. a,** Band dispersion of RG under applied displacement field; band gap $\tilde{\Delta}$ is masked by a bandwidth of $2\gamma_4\gamma_1/\gamma_0$ such that only the gap $\Delta = \tilde{\Delta} - 2\gamma_4\gamma_1/\gamma_0$ is visible in transport measurements. **b,** Calculated dependence of $\tilde{\Delta}$ and $\Delta$ on displacement field for $N = 9$ layers.